\documentclass[conference]{IEEEtran}
\IEEEoverridecommandlockouts

\newcommand{\IEEEAcceptedNotice}{%
To appear in the Proceedings of the 2026 IEEE Middle East Conference on Communications and Networking (MECOM 2026).\\
\copyright~2026 IEEE. Personal use of this material is permitted.  Permission from IEEE must be obtained for all other uses, in any current or future media, including reprinting/republishing this material for advertising or promotional purposes, creating new collective works, for resale or redistribution to servers or lists, or reuse of any copyrighted component of this work in other works.
}

\makeatletter
\def\ps@IEEEtitlepagestyle{%
  \def\@oddfoot{%
    \parbox[b]{0.96\textwidth}{%
      \centering
      \scriptsize
      \IEEEAcceptedNotice
    }%
  }%
  \def\@evenfoot{}%
}
\makeatother
\usepackage[utf8]{inputenc}
\usepackage{cite}
\usepackage{amsmath,amssymb,amsfonts}
\usepackage{algorithmic}
\usepackage{graphicx}
\usepackage{textcomp}
\usepackage{xcolor}
\usepackage{booktabs}
\usepackage{xurl}                  

\usepackage[caption=false,font=footnotesize]{subfig}

\def\BibTeX{{\rm B\kern-.05em{\sc i\kern-.025em b}\kern-.08em
    T\kern-.1667em\lower.7ex\hbox{E}\kern-.125emX}}
\begin{document}

\title{Evasion Attacks on Cost-Utility-Based Adversarial
Training for Online AutoML in IoT Networks}


\author{
    \IEEEauthorblockN{
        \IEEEauthorrefmark{1}\IEEEauthorrefmark{2}Chukwunonso Henry Nwokoye,
        \IEEEauthorrefmark{1}Wajiha Zaheer,
        \IEEEauthorrefmark{1}Khalil El-Khatib,
        and \IEEEauthorrefmark{1}Li Yang
    }
    \IEEEauthorblockA{
        \IEEEauthorrefmark{1}\textit{Faculty of Business and Information Technology,}
    \textit{Ontario Tech University,}
        Oshawa, Ontario, Canada \\
        \IEEEauthorrefmark{2}\textit{Department of Computer Science,}
    \textit{Alex Ekwueme Federal University,}
        Nigeria \\
        Emails: \{henrychukwunonso.nwokoye, wajiha.zaheer, khalil.el-khatib, li.yang\}@ontariotechu.ca
    }
}


\maketitle

\begin{abstract}
As IoT networks increasingly depend on machine learning for anomaly, malware, intrusion detection, and network monitoring, such systems have become attractive targets for evasion attacks. Evasion attacks pose a major security risk because an adversary intentionally modifies input data to mislead a trained model into producing incorrect predictions while evading detection. This study evaluates the impact of black-box evasion attacks on a cost-utility-based adversarial training defense strategy in an Online AutoML context for IoT networks. Specifically, evasion attacks were applied to online learners, including HT, LB, SRP, HAT, and ARF. By developing naive and adversarially trained (AT) versions of these online learners, we generated clean and adversarial accuracies for each model. The results show that the AT versions of LB and SRP performed best, achieving the highest adversarial accuracy (0.985) and high clean accuracy (0.993) at the highest cost budget of 1.00, with a maximum accuracy reduction of only 0.8\%. Finally, drift detection was conducted using the EDDM.

\end{abstract}


\section{Introduction} 


Machine learning (ML)-based intrusion detection systems (IDS) have emerged as an effective approach to identifying malicious activity in modern network environments \cite{Hiremath2025}. However, many traditional ML models are developed using static datasets and assume relatively stable data distributions \cite{yang2022iot}. In Internet of Things (IoT) environments, network traffic is generated as continuous data streams characterized by high volume, velocity, and variability \cite{IoTDataHandbook}. Changes in user behavior, device configurations, network conditions, and attack patterns can alter the underlying data distribution over time, leading to concept drift and subsequent degradation of model performance. Consequently, maintaining accurate and reliable intrusion detection in dynamic environments remains a significant challenge \cite{yang2022iot}. 

To address these limitations, online learning approaches have been proposed to incrementally update models as new data becomes available. Unlike batch learning methods that require access to complete datasets prior to training, online learning algorithms process data one sample at a time, enabling real-time adaptation to evolving environments while reducing memory requirements. Building upon these capabilities, Online Automated Machine Learning (Online AutoML) \cite{Zaheer2026} extends automation to tasks such as model selection, hyperparameter optimization, feature engineering, and model adaptation in streaming environments. By reducing the need for manual intervention while maintaining adaptability to changing data distributions, Online AutoML has emerged as a promising framework for autonomous intrusion detection in dynamic IoT networks \cite{yang2022iot}. Despite these advantages, the security of Online AutoML systems against adversarial machine learning attacks remains insufficiently explored, posing a critical vulnerability in modern network defense systems. Among these threats, evasion attacks pose the most severe and consistently effective risk to online ML systems.

Recent research demonstrates the serious impact of these attacks: reinforcement learning-based evasion methods achieve an average attack success rate of 94\% against flow classifiers \cite{Amoeba}, while advanced techniques such as EvadeRL have achieved 100\% evasion rates in later attack rounds against malware detection systems \cite{EvadeRL}. 

The real-time constraints of IoT environments add further complexity to adversarial robustness. Although such constraints may limit attack effectiveness, advanced evasion methods can still achieve up to 98\% of ideal clairvoyant attack performance \cite{joe2022onlineevasionattacksrecurrent}. In addition, adversarial perturbations may interact with concept drift in unpredictable ways, either amplifying or, in rare cases, reducing attack effectiveness \cite{NIDS}. This is especially concerning for Online AutoML systems, which continuously adapt their models, hyperparameters, and feature-selection strategies as new data arrive. Unlike static ML models with fixed decision boundaries, adaptive Online AutoML pipelines may incorporate adversarial inputs into their future optimization process, causing effects that extend beyond the immediate misclassification of individual samples. However, the impact of evasion attacks on continuously adapting Online AutoML-based IDS remains largely unexplored. Therefore, this study evaluates the robustness of Online AutoML-based IDS under black-box evasion attacks in streaming IoT environments.

This study evaluates the impact of black-box evasion attacks on an Online AutoML pipeline for streaming IoT intrusion detection. \footnote{Code and supplementary results are available at: \url{https://github.com/ChiNonsoHenry16/AutoML-AT-O}}. The learner pool consists of Hoeffding Tree (HT), Leveraging Bagging (LB), Streaming Random Patches (SRP), Adaptive Random Forest (ARF), and Hoeffding Adaptive Tree (HAT) \cite{yang2022iot}. The pipeline integrates automated preprocessing, imputation, normalization, feature engineering, online learner evaluation, and adversarial training (AT) under a cost-utility attack model. The models are evaluated on the public IoT Intrusion 2020 (IoTID20) dataset \cite{10.1007/978-3-030-47358-7_52}.



The main contributions of this paper are as follows:
\begin{enumerate}
\item We extend a streaming Online AutoML pipeline for IoT intrusion detection by incorporating cost-utility-aware AT.
\item We evaluate the robustness of five online learners, namely HT, LB, SRP, ARF, and HAT, under black-box evasion attacks with different cost budgets.
\item We compare naive and AT variants to quantify the trade-off between clean accuracy and adversarial accuracy.
\item We analyze concept-drift behavior under adversarial settings using EDDM and discuss how AT affects both robustness and online rolling accuracy stability.
\end{enumerate}

The remainder of this paper is organized as follows. Section II reviews related work on Online AutoML and adversarial robustness. Section III presents the methodology, including online learners, evasion attacks, AT, and drift detection. Sections IV and V report and discuss the results, and Section VI concludes the paper.

\section{Related Work}
\label{lit_rew}



Recent work on IoT analytics highlights the insufficiency of static AutoML pipelines, particularly due to concept drift in evolving network environments \cite{yang2022iot}. While Singh \textit{et al.} demonstrate AutoML's viability for intrusion detection via Bayesian optimization, their evaluation remains offline and excludes adversarial inputs \cite{Singh2022}. To manage streaming drift, Martindale \textit{et al.} show that heterogeneous online ensembles, specifically pairing Adaptive Random Forest with Hoeffding Adaptive Trees, balance drift recovery and accuracy but incur high computational costs that motivate adaptive configuration strategies for resource-constrained edge devices \cite{Martindale2020}.

OnMAR \cite{yang2022iot} introduces meta-learning for real-time AutoML, reusing prior designs unless an XGBoost predictor signals degradation, thus reducing runtime compared to genetic alternatives. However, its design assumes benign inputs, leaving it vulnerable in security-critical contexts. Concurrently, adversarial learning for tabular IoT traffic has moved beyond $L_p$-norm commonly used in the image domain. Kireev \textit{et al.} propose a cost-utility framework in which adversaries operate under bounded resource costs and heterogeneous gains, executing black-box targeted evasion attacks without poisoning the training data \cite{kireev2022adversarial}; this aligns with the black-box threat model described in prior work \cite{juuti2019making}. Their defense reveals a critical asymmetry: models hardened against utility-aware attackers generalize to cost-bounded threats, but not conversely.

Despite these advances, research on Network Intrusion Detection Systems (NIDS) robustness has not sufficiently examined cost-aware adversarial models in streaming Online AutoML pipelines. Existing Online AutoML studies mainly focus on model adaptation, runtime efficiency, and concept drift under benign data streams, while adversarial learning studies often assume static or offline learning settings. Therefore, the combined challenge of preserving robustness against cost-constrained evasion while maintaining adaptation to real-time concept drift remains unresolved.

Regarding adversarial knowledge, Kireev \textit{et al.} \cite{kireev2022adversarial} assume black-box access to the target classifier, where the adversary can issue queries using arbitrary examples and observe the prediction output $f(x)$. Therefore, the attack considered in this paper is a black-box evasion attack. It perturbs input samples at inference time to induce misclassification, but it does not modify the training data or poison the model. Considering the papers reviewed herein, we extend the Online AutoML pipeline proposed by Yang and Shami \cite{yang2022iot} by applying the cost-utility-aware AT strategy \cite{kireev2022adversarial}. We then evaluate the impact of evasion attacks on the improved pipeline and examine clean performance, adversarial robustness, and concept-drift behavior under adversarial conditions.

\section{Methodology}
\label{met}

\begin{figure}[htbp]
    \centering
    \includegraphics[width=8cm]{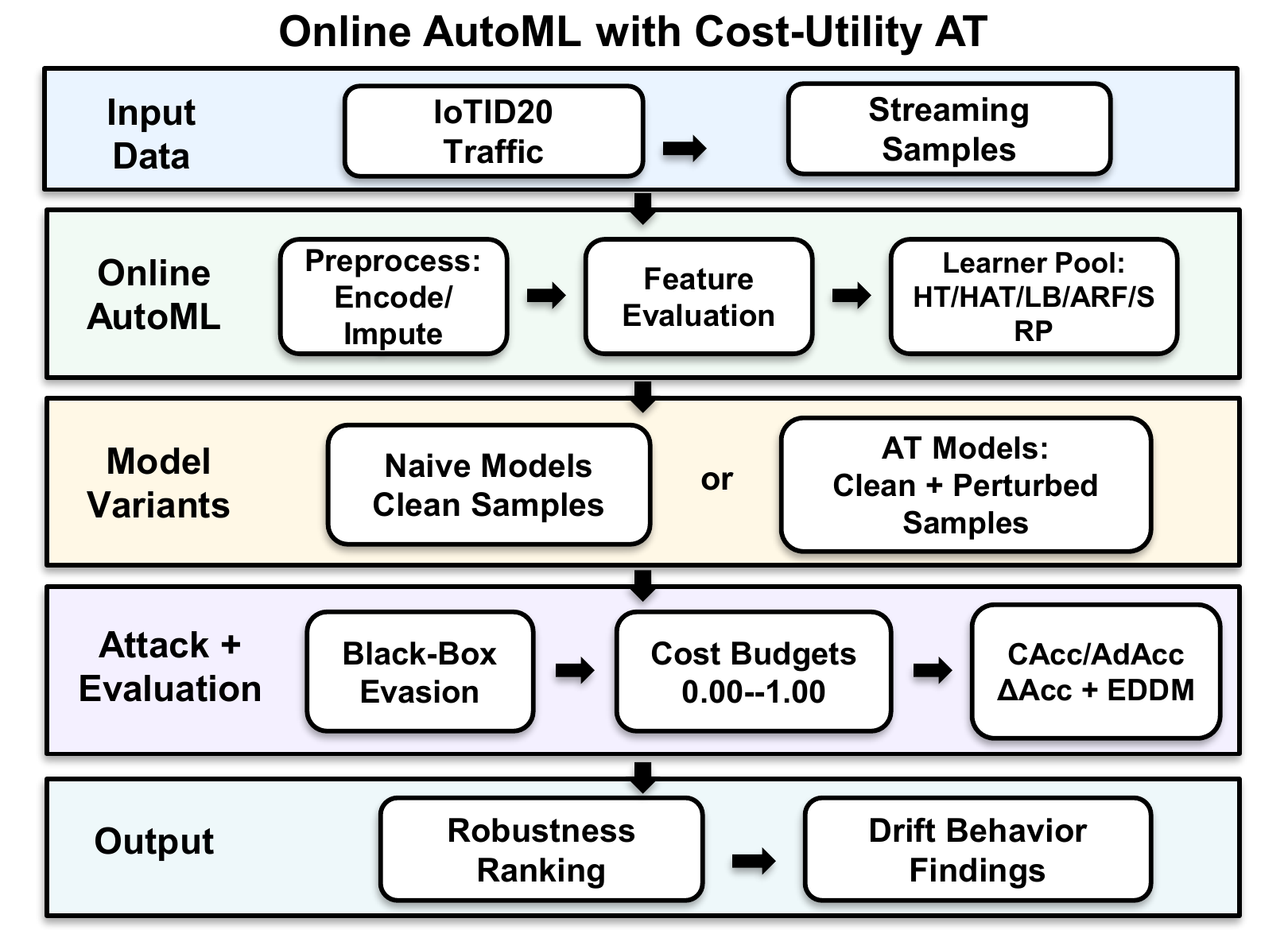}
    \caption{The overview of the proposed framework.}
    \label{fig:framework}
\end{figure}

We adopted the streaming AutoML pipeline proposed by Yang and Shami \cite{yang2022iot}. The Online AutoML pipeline \cite{yang2022iot, Zaheer2026} includes automated data preprocessing, auto-encoding, imputation, normalization, feature engineering, and online learning. In this study, categorical features were encoded, missing values were imputed, numerical features were normalized, and feature engineering was performed using information gain or Pearson correlation. The proposed framework is shown in Fig. \ref{fig:framework}. The study employed the IoTID20 dataset \cite{10.1007/978-3-030-47358-7_52}, and performance was measured using accuracy, precision, recall, and F1-score.

The online learning models used here include Hoeffding Tree (HT), Leveraging Bagging (LB), Streaming Random Patches (SRP), Adaptive Random Forest (ARF), and Hoeffding Adaptive Tree (HAT) \cite{yang2022iot}. These learners were selected because they represent different categories of online learning methods. HT provides a simple incremental tree-based baseline, while HAT extends HT by introducing adaptation to concept drift. LB represents online bagging-based ensemble learning, ARF combines adaptive trees with ensemble learning, and SRP increases diversity through random feature and instance subspaces. Therefore, the selected learners provide a diverse basis for evaluating whether AT affects different types of streaming models in similar or different ways.

\color{black}

For the generation of attacks and real-time training, all predictor variables incur unit modification costs, and numerical attributes are evaluated in dictionary order without a fixed-feature constraint. For every cost-effective attribute \(j\), the generator evaluates \(x_j \pm 0.2(u_j-l_j)\), constrained within the initialization-set limits \([l_j,u_j]\). Candidates are assessed based on label predictions and the actual label, employing misclassification loss \(\ell(f(x'),y)=\mathbf{1}\{f(x')\neq y\}\).

A modification is approved solely if it definitely enhances this loss; the whole expense corresponds to the count of permitted feature alterations. Generation employs a maximum of three predictive inquiries for each cost-effective numeric attribute \(3d\), without imposing an additional query limit. Clipping is mandated, although the validity of categorical/integer values, consistency among features, and the preservation of explicit usefulness are not enforced. Thus, its execution represents a cost-limited enhancement of the feature space instead of a complete utility-preserving alteration of traffic. Training consumes budget \(0.2d\). Every streamed clean prediction is documented prior to refreshing the learner with the clean record, so producing its equivalent with the revised model, and learning that equivalent. When the generator retrieves the initial record, the subsequent update repeats the clean instance.

The cost refers to the quantity of approved feature modifications, each incurring a single unit—not a financial expenditure. Utility would signify the maintenance of the assailant's desired advantage or operational capability. Nonetheless, validity tests for categorical/integer and inter-feature attributes are not executed. This approach is thus a cost-limited enhancement of feature space. 

A limited search was deemed necessary to mimic and maintain the computational feasibility of adversarial sample generation amid online learning, establishing a preliminary benchmark instead of a comprehensive assessment of potential assaults.

\color{black}

In the experimental setup, we developed naive and adversarially trained (AT) models for each of the online learners (HT, LB, ARF, SRP, and HAT). The resulting models are Naive-HT (Naive-HT), Adversarially Trained HT (AT-HT), Naive-LB (Naive-LB), Adversarially Trained LB (AT-LB), Naive-SRP (Naive-SRP), Adversarially Trained SRP (AT-SRP), Naive-ARF (Naive-ARF), Adversarially Trained ARF (AT-ARF), Naive-HAT (Naive-HAT), and Adversarially Trained HAT (AT-HAT).

Implementing both naive and AT model variants is important for assessing robustness against evasion attacks \cite{Yang2024}. Naive models were trained only on clean samples and served as baselines for standard streaming classification. In contrast, AT models were trained with adversarially perturbed samples generated under the cost-utility attack model. During AT, the adversarial samples were used together with the corresponding clean samples so that each learner could preserve standard detection performance while improving robustness to perturbed inputs. The considered attack is a black-box evasion attack: the adversary can query the classifier and observe its prediction output, but does not access model parameters or modify the training data. During evaluation, the attack perturbs test samples at inference time to induce misclassification while respecting a predefined cost budget. Clean and adversarial accuracies (CAcc and AdAcc) were then generated for naive and AT models under cost budgets from 0.00 to 1.00.

AT was selected because it is a direct and practical defense strategy against evasion attacks. Instead of only detecting adversarial inputs after deployment, AT exposes the model to adversarially modified samples during training, allowing the learner to adjust its decision boundary \cite{Yang2024}. This makes AT suitable for evaluating whether online learners can maintain reliable detection performance under cost-constrained adversarial manipulation.

\color{black}
The accuracy of online performance is assessed via a clean-input test-then-train methodology. Independently, the assessment of attack budgets halts the finalized trained models and compares unaltered and modified predictions for budgets \(B=bd\), where
\(b\in\{0,0.05,0.10,0.20,0.40,0.80,1.00\}\). This subsequent assessment reexamines the identical stream records previously utilized for learning. Thus, its accuracies reflect post-training resilience on previously encountered data.

\color{black}

For drift detection, we considered the Early Drift Detection Method (EDDM), Adaptive Windowing (ADWIN), and Drift Detection Method (DDM) \cite{yang2022iot}. DDM monitors changes in the online error rate, ADWIN uses an adaptive window to detect statistically significant distribution changes, and EDDM focuses on changes in the distance between classification errors, making it useful for gradual drift. While we envision drift detection as a trigger for adaptation mechanisms in real-time IoT settings, this study conducted separate assessments for EDDM, ADWIN, and DDM. Since EDDM produced clearer drift signals in our experiments, the reported drift figures focus on EDDM.

\section{Results}
\label{res}

To evaluate ML models in the context of adversarial threats, it is important to report both CAcc and AdAcc for naive and AT model variants. CAcc evaluates model performance on unaltered test data, indicating standard prediction capability. AdAcc assesses robustness against adversarial perturbations intended to trigger misclassification. This evaluation strategy is commonly used in adversarial robustness research \cite{goodfellow2014explaining,madry2017towards,ilyas2019adversarial}, where both clean and adversarial performance are needed to understand the trade-off between accuracy and robustness. Before the cost-budget analysis, we first equipped the Online AutoML pipeline with AT and evaluated it using accuracy metric.


\subsection{Impact of Evasion Attacks}

Here, we present CAcc, AdAcc, and their differences for the AT models under cost budgets from 0.00 to 1.00. At a cost budget of 0.00, CAcc = AdAcc for every model used in the study.

\begin{table}[t]
\centering
\caption{Comparison of CAcc \& AdAcc across AT models under different Cost Budgets}
\label{tab:combined_at_results}
\begin{tabular}{lcccc}
\hline
\textbf{Model} & \textbf{CAcc} & \textbf{AdAcc@0.05} &
\textbf{AdAcc@1.00} & \textbf{$\Delta$Acc@1.00} \\
\hline
AT-HT  & 0.987 & 0.987 & 0.963 & 0.024 \\
AT-LB  & 0.993 & 0.993 & 0.985 & 0.008 \\
AT-SRP & 0.993 & 0.993 & 0.985 & 0.008 \\
AT-HAT & 0.986 & 0.986 & 0.933 & 0.053 \\
AT-ARF & 0.994 & 0.994 & 0.957 & 0.037 \\
\hline
\end{tabular}
\end{table}


For the AT-HT model, CAcc consistently stayed at 98.7\% across every assessed budget, indicating steady performance on unperturbed data. Furthermore, the AdAcc equaled the CAcc at 0.00 and 0.05 cost budgets, indicating no performance degradation and illustrating that evasion attacks with minimal perturbation were unsuccessful against the model. With the budget cost rising to 0.10 and higher (1.00), the AdAcc had a slight decline from 98.7\% to 96.3\%, resulting in a $\Delta$Acc of 0.024. This decline then remained stable up to 1.00, suggesting that the attack effect saturated at a relatively small cost budget.




For the AT-LB model, the CAcc consistently stayed at 99.3\% across all cost budgets, indicating reliable prediction performance on unperturbed data. At budgets of 0.00 and 0.05, the AdAcc matched the CAcc (99.3\%), demonstrating no degradation of performance and suggesting that evasion attacks with minimal budgets of perturbation were ineffective. As the cost budget rose to 0.10 or higher, the AdAcc decreased to 98.5\%, corresponding to a $\Delta$Acc of 0.008. This decrease persisted throughout all elevated cost budgets up to 1.00, indicating that the evasion attack's effect reached saturation at a relatively modest perturbation cost budget and failed to induce further deterioration in model performance notwithstanding increased attack costs.




For the AT-SRP model, the CAcc consistently stayed at 99.3\% across various assessed cost budgets, indicating steady classification performance for unperturbed data. Furthermore, the AdAcc matched the CAcc at 0.00 and 0.05 cost budgets, indicating no reduction in accuracy and illustrating that evasion attacks with minimal budgets of perturbation had no apparent impact on the AT-SRP model. As the budget cost rose to 0.10 and above, the AdAcc had a slight decline to 98.5\%, corresponding to a $\Delta$Acc of 0.008. This decline remained stable up to 1.00, suggesting that additional attack resources did not further reduce performance.




For the AT-HAT model, the CAcc consistently measured 98.6\% across every assessed budget, indicating steady performance on unperturbed data. At 0.00 and 0.05 cost budgets, the AdAcc equaled the CAcc, demonstrating no performance decrease and suggesting that evasion attacks with minimal budgets of perturbation were unsuccessful against the model. As the attack cost budget rose to 0.10 or higher (0.10 - 1.00), the AdAcc dropped from 98.6\% to 93.3\%, resulting in a $\Delta$Acc of 0.053. This decrease was consistent across all elevated cost budgets from 0.0 to 1.00, indicating that the evasion attack achieved its peak effect at a relatively minimal perturbation budget and that subsequent attacks did not significantly impair the performance of the model.




For the AT-ARF model, the CAcc consistently remained at 99.4\% across every assessed cost budget, indicating robust performance on unperturbed data. At 0.00 and 0.05 cost budgets, the AdAcc matched the CAcc, indicating no decrease in accuracy and illustrating that evasion attacks with minimal budgets of perturbation were unsuccessful against the ARF classifier. As the cost budget rose to 0.10 and higher (0.10 - 1.00), the AdAcc diminished from 99.4\% to 95.7\%, resulting in a $\Delta$Acc of 0.037. This decline persisted uniformly over all elevated cost budgets from 0.00 to 1.00, signifying that the evasion attack achieved its peak efficacy at a relatively small budget of perturbation and that subsequent attack resources did not significantly undermine the model's performance.



\begin{figure*}[!t]
\centering

\subfloat[Naive-LB]{
    \includegraphics[width=0.48\textwidth]{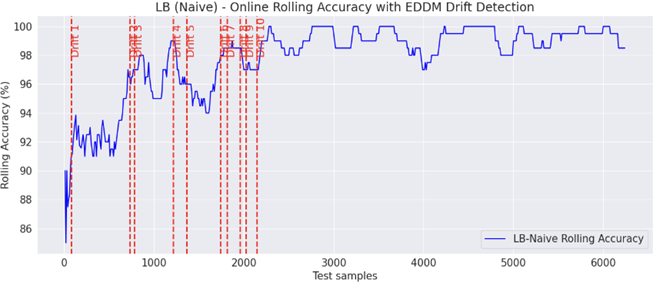}
    \label{fig:nlb-eddm}
}
\hfil
\subfloat[AT-LB]{
    \includegraphics[width=0.48\textwidth]{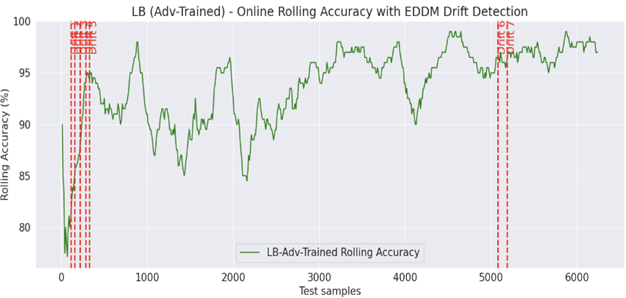}
    \label{fig:atlb-eddm}
}

\vspace{1mm}

\subfloat[Naive-SRP]{
    \includegraphics[width=0.48\textwidth]{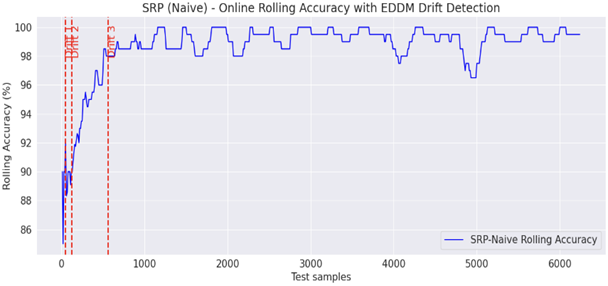}
    \label{fig:nsrp-eddm}
}
\hfil
\subfloat[AT-SRP]{
    \includegraphics[width=0.48\textwidth]{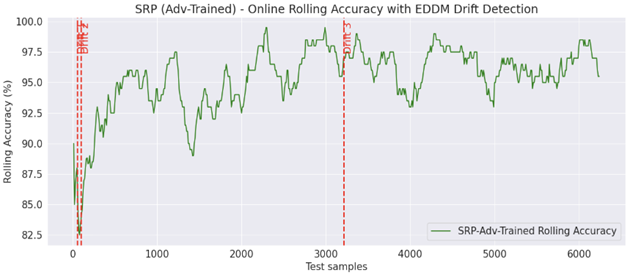}
    \label{fig:atsrp-eddm}
}

\caption{EDDM-based drift detection results for the naive and adversarially trained LB and SRP models.}
\label{fig:eddm-drift}
\end{figure*}


\begin{figure}[htbp]
    \centering
    \includegraphics[width=9cm]{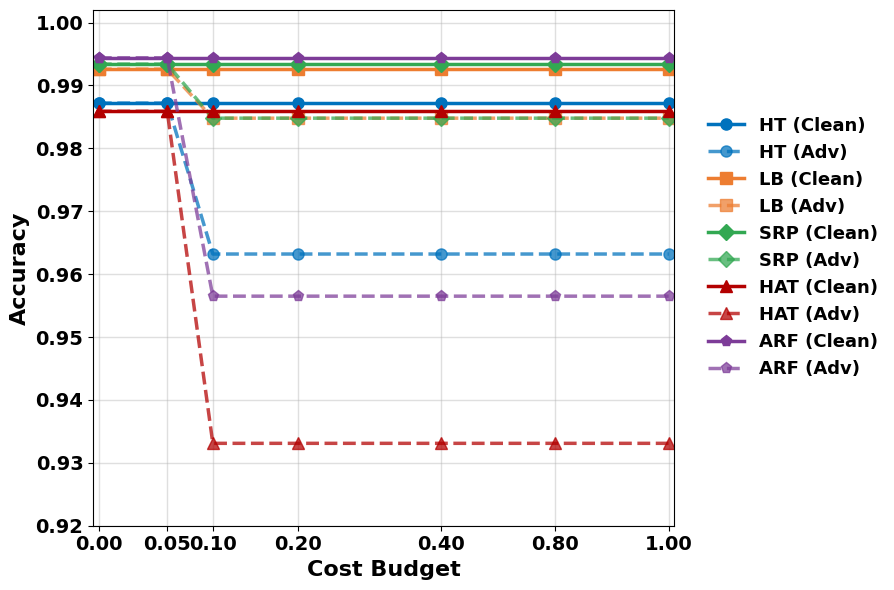}
    \caption{Clean and adversarial accuracies under different cost budgets.}
    \label{fig:acc-n-cb}
\end{figure}


\color{black}

CAcc remains unchanged since both the model and the unaltered inputs are constant. However, AdAcc matches CAcc when no alterations are affordable (or below the unit-cost threshold). When a modification leads to misclassification, the binary loss attains its peak value of one, inhibiting any more acceptable alterations. Isolated ineffective modifications are eliminated instead of merged. As a result, for deterministic forecasts, every evaluated budget from $(b=0.10)$ onward yields the same altered inputs. This plateau indicates a limited search, rather than a proven resilience against more robust multi-feature evasion attacks. The investigation offers a computationally efficient foundation for online learning, rather than a comprehensive evaluation of potential attacks.

\color{black}

\subsection{Drift Detection using EDDM}

In streaming (online) learning contexts, the fundamental distribution of data may change over time, a process referred to as concept drift. Identifying such drift is essential, since it enables ML models to adjust to varying and ever-evolving circumstances and sustain superior predicted performance. Various drift detection techniques have been introduced, with the Early Drift Detection Method (EDDM) exhibiting notable efficacy in detecting changes by observing "increases in the distance between classification errors"~\cite{baena2006eddm}. Here, we employed EDDM to rigorously examine and contrast the drift sensitivity attributes of both naive and AT models in streaming IoT data environments. The subsequent findings underscore variations in the frequency of drift detection and the correlation between drift occurrences and the stability of model classification accuracy. It is noteworthy that we applied other drift detection methods, such as ADWIN and DDM; however, EDDM showed more drift occurrences for the models.

Fig.~\ref{fig:eddm-drift} compares online rolling accuracy (ORA) and EDDM-detected drifts for the naive and adversarially trained LB and SRP models. Subfigures (a) and (b) show the results for Naive-LB and AT-LB, respectively. The Naive-LB model attains higher and more consistent ORA, although it shows a greater number of drift detections, mostly during the early stages of the assessment period. Conversely, the AT-LB model exhibits lower and more variable ORA, with fewer identified drifts.

Subfigures \ref{fig:nsrp-eddm} and \ref{fig:atsrp-eddm} show the EDDM-supported drift detection results for Naive-SRP and AT-SRP, respectively. The Naive-SRP model attains higher and more consistent rolling accuracy, with three identified drift occurrences during the first 500 test samples. In contrast, the AT-SRP model exhibits lower and more fluctuating accuracy throughout the stream, with only two detected drifts. These findings suggest that AT may reduce the frequency of detected drifts, but it does not necessarily improve online rolling accuracy stability.

\color{black}
EDDM values diminish for HT (from 11 to 6) and LB (from 10 to 7), while remaining constant for SRP and ARF. ADWIN values drop for HT and LB, stay constant for SRP, and rise for ARF; DDM counts escalate for LB, SRP, and ARF. However, these varying responses do not entirely demonstrate enhanced resilience to concept drift. The diminished and inconsistent ORA of AT-LB and AT-SRP warns against the assumption that a reduced number of alerts signifies superior adaptation. 


\color{black}
\section{Discussion}
\label{sec:discussion}

Our experiments show that the adversarially trained versions of Leveraging Bagging (AT-LB) and Streaming Random Patches (AT-SRP) performed best, with the highest AdAcc (0.985) and very high CAcc (0.993) at the highest cost budget of 1.00. The difference between CAcc and AdAcc is minimal for Leveraging Bagging (AT) and SRP (AT), at only 0.008. The other online learners exhibit less AdAcc across all higher cost budgets (0.10 to 1.00). 
The AT-LB and AT-SRP models exhibit significant resilience to cost-based black-box evasion attacks, sustaining accuracies above 98\% alongside a maximum accuracy decline of just 0.8\%. The findings indicate that the AT defense strategy significantly improves the classifiers' robustness, enabling them to maintain high performance in more severe attack scenarios. This trend is visually illustrated in Fig. \ref{fig:acc-n-cb}.

While AT implementations of LB and SRP performed better, we sought to determine the best model between them. Furthermore, an examination of naive versions (Tables \ref{tab:n-lb} and \ref{tab:n-srp}) of LB and SRP reveals notable disparities in their resilience to adversarial alterations. LB attains somewhat superior CAcc compared to SRP (\textbf{\textit{0.996}} for LB against \textbf{\textit{0.994}} for SRP). Nonetheless, LB exhibits a much greater reduction in AdAcc as the cost budget grows. For budgets of 0.10 and higher, LB's AdAcc decreases to 0.834, resulting in a $\Delta$Acc (CAcc–AdAcc difference) of \textbf{\textit{0.162}}, while SRP's AdAcc declines to 0.914 with a reduced $\Delta$Acc of \textbf{\textit{0.080}}. This suggests that SRP is inherently more resilient to adversarial samples even without AT, likely due to its random feature and instance subspaces.

\begin{table}[h]
    \centering
    \caption{Naive-LB: Clean and Adversarial Accuracies}
    \label{tab:n-lb}
    \begin{tabular}{cccc}
        \toprule
        \textbf{Cost budget} & \textbf{CAcc} & \textbf{AdAcc} & \textbf{$\Delta$Acc} \\
        \midrule
        0.00 & 0.996 & 0.996 & 0.000 \\
        0.05 & 0.996 & 0.996 & 0.000 \\ \hline
        0.80 & 0.996 & 0.834 & 0.162 \\ \hline
        \textbf{1.00} & \textbf{0.996} & \textbf{0.834} & \textbf{0.162 (16.2\%)} \\
        \bottomrule
    \end{tabular}
\end{table}

\begin{table}[h]
    \centering
    \caption{Naive-SRP: CAcc, AdAcc, and their Difference}
    \label{tab:n-srp}
    \begin{tabular}{cccc}
        \toprule
       \textbf{Cost budget} & \textbf{CAcc} & \textbf{AdAcc} & \textbf{$\Delta$Acc} \\
        \midrule
        0.00 & 0.994 & 0.994 & 0.000 \\
        0.05 & 0.994 & 0.994 & 0.000 \\\hline
        0.80 & 0.994 & 0.914 & 0.080 \\ \hline
        \textbf{1.00} & \textbf{0.994} & \textbf{0.914} & \textbf{0.080 (8\%)} \\
        \bottomrule
    \end{tabular}
\end{table}

\section{Conclusion}
\label{con}

This study examined the impact of black-box evasion attacks on an Online AutoML pipeline equipped with a cost-utility-based AT defense strategy. Reporting both clean accuracy (CAcc) and adversarial accuracy (AdAcc) enabled a direct evaluation of each model's trade-off between standard performance and robustness under attack. The results show that the adversarially trained versions of Leveraging Bagging and Streaming Random Patches performed best, achieving the highest AdAcc (0.985) and very high CAcc (0.993) at the highest cost budget of 1.00, with an accuracy reduction of only 0.8\%. Among the naive implementations, SRP performed better than LB under adversarial perturbations, suggesting that SRP is preferable when resilience without AT is important. Overall, naive models often achieve high clean accuracy but may suffer larger performance drops under adversarial samples, while AT models improve robustness by reducing the gap between clean and adversarial performance. Finally, EDDM was used to detect drift in the naive and AT versions of the best-performing models. Future work will extend the evaluation to additional datasets, repeated runs, and runtime/resource analysis.

\section*{Acknowledgment}

This project was made possible in part through the support of the National Cybersecurity Consortium and the Government of Canada (CSIN).

\bibliographystyle{ieeetr}
\bibliography{references}

\vspace{12pt}
\end{document}